\documentclass[%
 reprint,
superscriptaddress,
 amsmath,amssymb,
 aps,
pra,
]{revtex4-2}

\usepackage{graphicx}
\usepackage{dcolumn}
\usepackage{bm}
\usepackage{xcolor}
\usepackage{hyperref}
\usepackage{comment}
\hypersetup{
    colorlinks=true,
    linkcolor=blue,
    citecolor=blue,
    filecolor=blue,
    urlcolor=blue
}
\begin{document}

\preprint{APS/123-QED}

\title{Resolving magnetic-sublevel structure in Rydberg Autler-Townes spectra with arbitrary RF polarization
}
\author{Noah~Schlossberger}
\affiliation{National Institute of Standards and Technology, Boulder, Colorado 80305, USA}
\email[]{noah.schlossberger@nist.gov}
\author{Rajavardhan~Talashila}
\affiliation{National Institute of Standards and Technology, Boulder, Colorado 80305, USA}
\affiliation{Department of Physics, University of Colorado, Boulder, Colorado 80309, USA}
\author{Stone~B.~Oliver}
\affiliation{National Institute of Standards and Technology, Boulder, Colorado 80305, USA}
\affiliation{Department of Physics, University of Colorado, Boulder, Colorado 80309, USA}
\author{Nikunjkumar~Prajapati}
\affiliation{National Institute of Standards and Technology, Boulder, Colorado 80305, USA}
\author{William~J.~Watterson}
\affiliation{National Institute of Standards and Technology, Boulder, Colorado 80305, USA}
 \author{Christopher~L.~Holloway}
\affiliation{National Institute of Standards and Technology, Boulder, Colorado 80305, USA}

\date{\today}

\begin{abstract}
We investigate the role of magnetic sublevels in Autler–Townes spectra of Rydberg atoms driven by radio-frequency (RF) fields with arbitrary polarization. While conventional treatments predict two symmetric sidebands from independent $m_J$ transitions, experiments have reported additional unexplained spectral features. We show that these arise from elliptical RF polarization, which coherently couples multiple magnetic sublevels and requires a full multi-level treatment.
We develop and diagonalize a Hamiltonian including all coupled $m_J$ sublevels, predicting polarization-dependent degeneracies that produce two, three, or four resolved peaks. Using long-wavelength transitions and an anechoic environment we realize homogeneous RF fields that for the first time enable  complete resolution of the $m_J$-dependent dressed states. We observe excellent agreement with theory as the RF ellipticity is varied.
These results demonstrate that RF polarization fundamentally modifies Autler–Townes spectra and provide a consistent framework for interpreting magnetic-sublevel structure, with implications for Rydberg-based RF electrometry and polarimetry.

\end{abstract}

\maketitle


\section{Introduction}
Autler-Townes splitting in Rydberg states of alkali atoms has been demonstrated as a sensitive, SI-traceable form of radio-frequency (RF) electrometry \cite{Schlossberger2024,9374680}. It has been shown that angular momentum plays a role in the shape of these spectra, and this effect has been leveraged to use Rydberg sensors as RF polarimeters \cite{PhysRevLett.111.063001, 10.1063/5.0285778, cloutman2026rydbergatomicpolarimetryradiofrequency}. The interpretation of these angular momentum effects has important implications for the traceability of these sensors \cite{9363580}. Theoretical treatments \cite{cloutman2026rydbergatomicpolarimetryradiofrequency} of the angular momenta predict that a $D\rightarrow P$ transition should split into two peaks on either side, in addition to the central unsplit peak. These peaks correspond to individual $m_J$ levels.  However, in previous measurements aiming to resolve angular-momentum-dependent behavior \cite{10.1063/5.0173545,cloutman2026rydbergatomicpolarimetryradiofrequency}, field inhomogeneity broadened the dressed states beyond their energy separation. Experiments that applied enough field strength to observe structure related to magnetic sublevels \cite{PhysRevA.109.L021702, 10.1063/5.0161213,10.1063/5.0088532} showed more than the two peaks predicted, and their relative locations did not agree with theory. Here we show that the observed additional peaks are due to elliptical polarization of the RF field (the presence of scatterers prevents pure RF polarization in many environments). We derive theoretical Autler-Townes spectra for arbitrary polarizations, and measure the spectra experimentally. To allow for resolution of the $m_J$ peaks, we apply RF with a wavelength long compared to the vapor cell to ensure spatial homogeneity \cite{PhysRevApplied.4.044015} and place the vapor cell in an anechoic chamber to allow for pure RF polarization. With this, we apply a large enough field to separate the $m_J$ structure and study the resulting spectra as a function of RF polarization. While our study focuses on a $D_{5/2} \rightarrow P_{3/2}$ transition, the analysis can easily be extended to other fine structure states.
\section{Theory}
\begin{figure}
    \includegraphics[scale = .9]{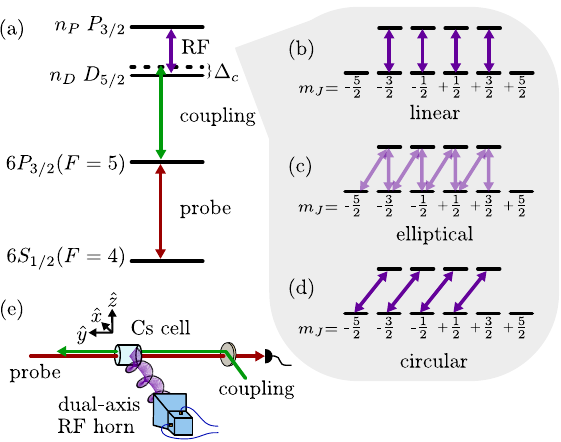}
    \caption{(a) The energy level diagram for the RF field sensing. The coupling of the two Rydberg states' $m_J$ subspaces are shown for  (b) linear, (c) elliptical, and (d) circular polarizations of the RF field. (e) The experimental layout and definition of the coordinate system.}
    \label{fig:EDL}
\end{figure}
The system is shown in Fig. \ref{fig:EDL}. We consider RF radiation resonant between two Rydberg states $|n_D D_{5/2}\rangle$ and $|n_PP_{3/2}\rangle$. The resonant RF radiation dresses the energies of these states \cite{y8j4-lsbr}. We then read out the energy of these dressed states using two-photon electromagnetically induced transparency (EIT), following the experimental scheme of Ref. \cite{Sedlacek2012}.

Pure linear (Fig. \ref{fig:EDL}b) and circular (Fig. \ref{fig:EDL}d) RF polarizations independently couple $m_J$ levels, and thus can be thought of as independent systems, whereby each $m_J$ is dressed and probed by the lasers separately with no coherent interactions. However, for elliptical polarization (Fig. \ref{fig:EDL}c), the effect of each $m_J$ are no longer independent as the RF coherently couples nearly all the $m_J$ states together. Because of this, the spectral peaks can not be treated as resulting from individual transitions with separate dipole moments. Instead, we must consider all the coherent couplings in a Hamiltonian which accounts for all magnetic sublevels and couplings, with the spectral peaks corresponding to eigenvalues of this Hamiltonian.

\begin{figure}
\includegraphics[scale = .9]{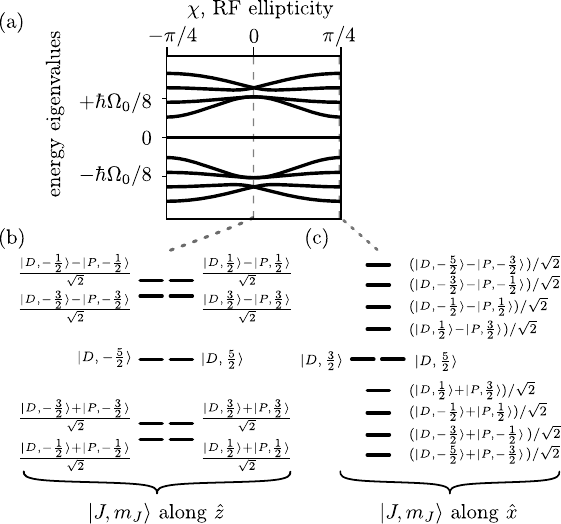}
\caption{(a) The energy eigenvalues as a function of the ellipticity $\chi$. (b) An energy level diagram of the dressed eigenstates for linearly polarized RF $(\chi = 0)$ in the basis along $\hat z$. (c) The dressed eigenstates for circularly polarized RF $(\chi = \pi/4)$ in the basis along $\hat x$.}
\label{fig:mJ_makeup}
\end{figure}

We write the RF electric field in terms of a complex phasor amplitude
\begin{equation}
\mathbf{E}_{\rm RF}(t)
=
\Re\!\left[
\tilde{\mathbf E}_{\rm RF}\,e^{-i\omega t}
\right]\,\,\, .
\end{equation}
Consider the Hamiltonian in the basis of all $m_J$ sublevels of the $D_{5/2}$ and $P_{3/2}$ Rydberg states coupled by the RF field. If we choose the quantization axis $z$ to be aligned with linear polarization of an RF field (consistent with the coordinate system defined in Fig. \ref{fig:EDL}e), we can write the complex RF field as
\begin{equation}
\tilde{\mathbf E}_{\rm RF}
=
E_0
\left(
\cos\chi\,\hat{\mathbf z}
+
i\sin\chi\,\hat{\mathbf y}
\right),
\label{eq:circularpolarizationdef}
\end{equation}
where $\chi$ parametrizes the ellipticity of the RF polarization. Here, $\chi = 0$ corresponds to linear polarization along $z$, which will drive $\pi$ transitions. In this basis, the spherical components of the field are
\begin{equation}
\epsilon_0 = \cos\chi,
\qquad
\epsilon_{+} = -\frac{1}{\sqrt{2}}\sin\chi,
\qquad
\epsilon_{-} = -\frac{1}{\sqrt{2}}\sin\chi .
\end{equation}
In this convention $\epsilon_0$ drives $\pi$ transitions ($\Delta m_J=0$), while $\epsilon_\pm$ drive $\sigma^\pm$ transitions ($\Delta m_J=\pm1$). In the ordered $|J, m_J\rangle$ basis:
\small
\begin{multline}
\{
|D,-\tfrac52\rangle,\,
|D,-\tfrac32\rangle,\,
|D,-\tfrac12\rangle,\,
|D,+\tfrac12\rangle,\,
|D,+\tfrac32\rangle,\,\\
|D,+\tfrac52\rangle,\,
|P,-\tfrac32\rangle,\,
|P,-\tfrac12\rangle,\,
|P,+\tfrac12\rangle,\,
|P,+\tfrac32\rangle
\},
\end{multline}\normalsize
using the appropriate Wigner 3-$j$ symbols \cite{edmonds1996angular}, the rotating-wave Hamiltonian (for zero detuning) takes the form
\begin{widetext}
\begin{equation}
H_{\rm RWA}
=
\frac{\hbar\Omega_0}{2}
\begin{pmatrix}
0&0&0&0&0&0&
\frac{1}{\sqrt6}\epsilon_{+}&0&0&0\\
0&0&0&0&0&0&
-\frac{1}{\sqrt{15}}\epsilon_{0}&
\frac{1}{\sqrt{10}}\epsilon_{+}&0&0\\
0&0&0&0&0&0&
\frac{1}{\sqrt{60}}\epsilon_{-}&
-\frac{1}{\sqrt{10}}\epsilon_{0}&
\frac{1}{\sqrt{20}}\epsilon_{+}&0\\
0&0&0&0&0&0&
0&
\frac{1}{\sqrt{20}}\epsilon_{-}&
-\frac{1}{\sqrt{10}}\epsilon_{0}&
\frac{1}{\sqrt{60}}\epsilon_{+}\\
0&0&0&0&0&0&
0&0&
\frac{1}{\sqrt{10}}\epsilon_{-}&
-\frac{1}{\sqrt{15}}\epsilon_{0}\\
0&0&0&0&0&0&
0&0&0&
\frac{1}{\sqrt6}\epsilon_{-}\\
\frac{1}{\sqrt6}\epsilon_{+}^{*}&
-\frac{1}{\sqrt{15}}\epsilon_{0}^{*}&
\frac{1}{\sqrt{60}}\epsilon_{-}^{*}&
0&0&0&
0&0&0&0\\
0&
\frac{1}{\sqrt{10}}\epsilon_{+}^{*}&
-\frac{1}{\sqrt{10}}\epsilon_{0}^{*}&
\frac{1}{\sqrt{20}}\epsilon_{-}^{*}&
0&0&
0&0&0&0\\
0&0&
\frac{1}{\sqrt{20}}\epsilon_{+}^{*}&
-\frac{1}{\sqrt{10}}\epsilon_{0}^{*}&
\frac{1}{\sqrt{10}}\epsilon_{-}^{*}&0&
0&0&0&0\\
0&0&0&
\frac{1}{\sqrt{60}}\epsilon_{+}^{*}&
-\frac{1}{\sqrt{15}}\epsilon_{0}^{*}&
\frac{1}{\sqrt6}\epsilon_{-}^{*}&
0&0&0&0
\end{pmatrix},
\label{eq:ellipticHamiltonian}
\end{equation}
\end{widetext}
where $\Omega_0$ is the Rabi rate associated with the reduced dipole matrix element $\mu_0$:
\begin{equation}
\Omega_0=\frac{\mu_0 E_0}{\hbar} = \frac{\langle n_P P_{3/2}\|\mu\|n_D D_{5/2}\rangle E_0}{\hbar}.
\end{equation}
The eigenvalues of this Hamiltonian are shown as a function of the ellipticity $\chi$ in Fig. \ref{fig:mJ_makeup}a. 
Because there are ten states involved in the coherent coupling, there will be ten dressed states with some degeneracy. The distribution of these states are shown for linearly polarized RF in Fig. \ref{fig:mJ_makeup}b and for circularly polarized RF in Fig. \ref{fig:mJ_makeup}c. For simplicity, the eigenstates are given in the quantization axis for which $m_J$ is a good quantum number.

We can note a few things from the theoretical result. The degeneracy of the dressed eigenstates changes for different $\chi$, meaning the polarization will affect the number of observed spectroscopic peaks. Furthermore, for pure linear and circular polarizations, each eigenstate is a simple superposition of the two states corresponding to each allowed coupling of the RF.

\section{Measurement}
We confirm the energy spectra predicted by Eq. \ref{eq:ellipticHamiltonian} with spectroscopic measurements using the EIT readout scheme from Fig. \ref{fig:EDL}e. To achieve resolution of the dressed eigenstates, we must ensure field homogeneity. We pick a set of Rydberg states $n_D = 65, n_P = 66$ corresponding to an RF transition wavelength ($\lambda=$120~cm) which is long compared to the vapor cell in order to minimize internal reflections. In addition, we perform the measurement in an anechoic chamber lined with RF-absorbing foam. With this and an RF amplifier to apply fields on the order of V/m, it becomes possible to separate the individual dressed eigenstates by more than their linewidth.

\begin{figure}
    \includegraphics[scale = .9]{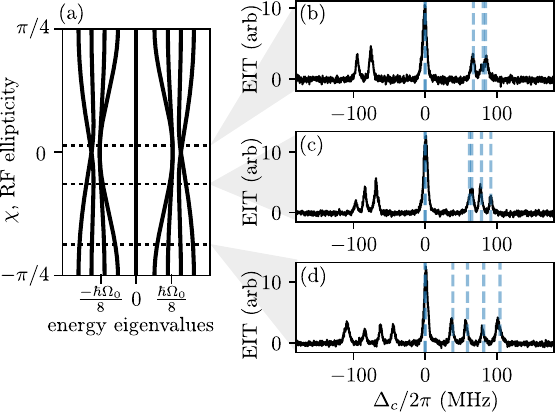}
    \caption{(a) Eigenvalues of the Hamiltonian from Eq. \ref{eq:ellipticHamiltonian} as a function of the RF ellipticity $\chi$. (b-d) Experimental spectroscopy of the Autler-Townes splitting at various ellipticities giving rise to (b) two, (c) three, and (d) four dressed peaks at a field strength of $E_0=5$~V/m. The predicted peak locations from the eigenvalues of Eq. \ref{eq:ellipticHamiltonian} are overlayed as dashed lines. The lasers are linearly polarized along $\hat x$ in (b) and (c) and along $\hat z$ in (d) to populate all $m_J$.}
    \label{fig:measurements_threepol}
\end{figure}

We lock the probe on resonance and scan the coupling laser over an angular-frequency detuning $\Delta_c$ from resonance (see Fig. \ref{fig:EDL}a). The measured spectra are shown in Fig. \ref{fig:measurements_threepol}.
At $\chi = 0$, where the RF polarization is purely linear, the eigenstates are maximally degenerate, leading to only two peaks on either side of the spectrum. However, the degeneracy of these peaks is lifted at impure polarizations. Spectra at various ellipticities are measured in Fig. \ref{fig:measurements_threepol}b-d. The apparent number of peaks varies from two to four with the ellipticity of the RF field.

\begin{figure}
\includegraphics[scale = .9]{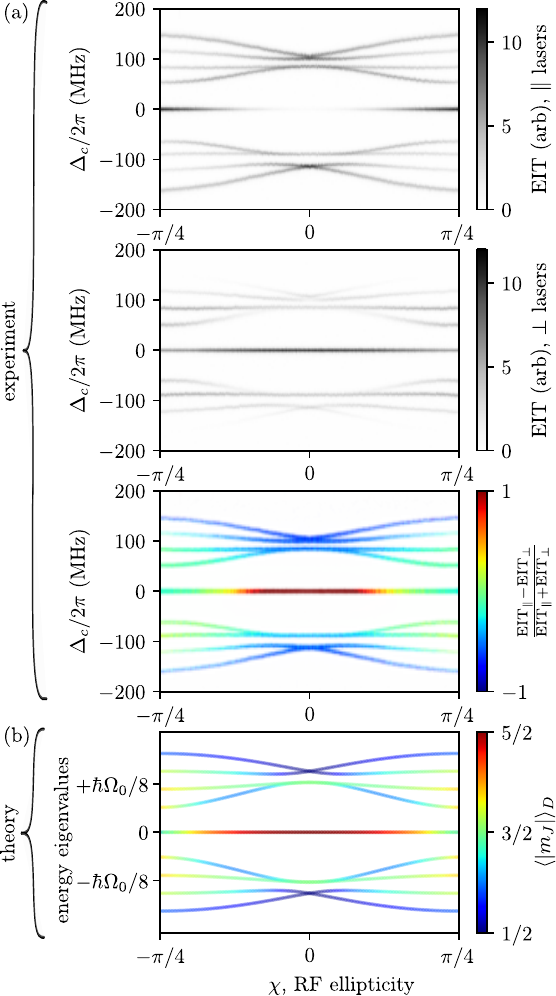}
\caption{a) Measured spectra of the $65D_{5/2} \rightarrow 66P_{3/2}$ transition at $E_0$ = 7~V/m scanning the ellipticity with the dual axis horn. The spectra is measured with the lasers parallel to $\hat z$ (top), and perpendicular to $\hat z$ (middle). We then take the normalized difference of the two spectra (bottom) to qualitatively approximate the angular momentum magnitude $|m_J|$ along $\hat z$. An alpha mask is applied based off of the sum of the spectra $\text{EIT}_\parallel + \text{EIT}_\perp$. b) The eigenvalues of Eq. \ref{eq:ellipticHamiltonian} are plotted with a colormap calculated from the expectation value of $|m_J|$ along $\hat z$ in the $D$ state (that which the lasers probe).}
\label{fig:m_Jcolorplot}
\end{figure}

To control the ellipticity with our dual-axis horn, we set the $\hat y$ component to be 90 degrees out of phase with the $\hat z$ component. We then scan the relative powers $P_z$ and $P_y$ of the two RF channels to realize the field defined in Eq. \ref{eq:circularpolarizationdef}:
\begin{equation}
        P_{z} = P_0 \cos^2(\chi),
        \qquad
        P_{y} = P_0  \sin^2(\chi),
\end{equation}
where $P_0$ is the total power sent to the RF horn. The measured spectra are shown as we scan $\chi$ continuously in Fig. \ref{fig:m_Jcolorplot}a. 

By repeating the measurement at two laser polarizations and comparing the spectra, we can perform a crude method of quantum state tomography to measure the $|m_J|$ of each peak. While the eigen-energies are independent of the basis/coordinate system, the basis in which $m_J$ is a good quantum number can rotate with the RF polarization. However, because the linearly polarized lasers have a coordinate system that does not change, they will measure $m_J$ along a consistent direction. Consider $\hat z$ to be the quantization axis (that in which Eq. \ref{eq:ellipticHamiltonian} is defined). The polarization of the lasers can be used to preferentially populate different $m_J$ states along $\hat z$. When the lasers are linearly polarized in $\hat z$, they preferentially populate and measure low $|m_J|$ states because they drive $\pi$ transitions whose transition dipole moments are stronger going to low $|m_J|$. When the lasers are pointed in $\hat x$, they drive $\sigma^+$ and $\sigma^-$ transitions in $\hat z$, which have larger dipole moments going to larger $|m_J|$. Therefore, the relative difference in the amplitudes of the spectra is a good qualitative estimation of the angular momentum magnitude $|m_J|$.
In Fig. \ref{fig:m_Jcolorplot}, we measure the spectra as a function of the RF ellipticity, which we set via the relative amplitudes of out-of-phase signals on the dual axis horn (consistent with Eq. \ref{eq:circularpolarizationdef}). We then take the difference of the measured spectra to approximate the angular momentum along $\hat z$. We then compare this to the expected $|m_J|$ in the D state for each eigenstate (Fig. \ref{fig:m_Jcolorplot}b), calculated as
\begin{equation}
    \langle |m_J|\rangle _D =  \frac{\sum_{m_J} |m_J| \langle D, m_J |\psi\rangle }{\sum_{m_J} \langle D, m_J | \psi \rangle }.
\end{equation}
where $\psi$ is the RF-dressed eigenstate.
\section{Pure RF polarization}
\begin{figure}
    \includegraphics[scale = .9]{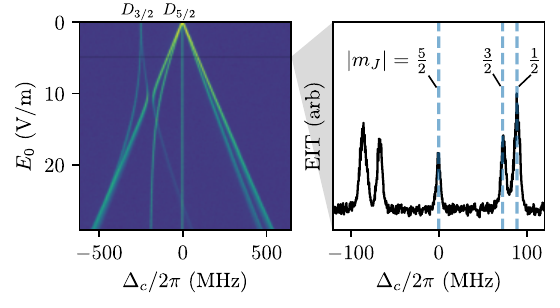}
    \caption{Measured Autler-Townes spectra of the 2.5~GHz $65D_{5/2} \leftrightarrow 66P_{3/2}$ transition with nearly pure linear RF polarization. Left: Autler-Townes spectra varying the RF field strength. Right: an individual spectrum at an $E_0$ of 4.9~V/m. The predicted peak locations from Eq. \ref{eq:pi_splittings} are overlayed as dashed lines.  In the plot on the right, we reduce the $\Delta_c$-axis scale to only highlight the D$_{5/2}$ states.}
    \label{fig:pipolarized}
\end{figure}

In general, linearly polarized RF radiation is difficult to obtain in the laboratory without special treatment due to scattering of the RF. However, theory and interpretation of spectra can be much simpler for pure linear polarization. Because it does not couple disparate $m_J$ states together, we can treat each $m_J$ ladder as an independent two-level subsystem. Each $|m_J|$ will have a different transition dipole matrix element for the same RF coupling and will therefore split a different amount. Looking at the couplings in Fig. \ref{fig:EDL}b, we can see that there are three distinct values of $|m_J|$. The highest value $|m_J|$ will not split, as there is no $|m_J| = 5/2$ in the $P$ state to couple to. The other two couplings are given by angular momentum algebra via the appropriate Wigner 3-$j$ symbol \cite{edmonds1996angular}:
\begin{equation}
   \Omega = \frac{E_0}{\hbar} \cdot \begin{cases}
       0, & |m_J| = 5/2\\
       \frac{1}{\sqrt{15}} \mu_0, & |m_J| = 3/2\\
       \frac{1}{\sqrt{10}} \mu_0, & |m_J| = 1/2
   \end{cases}\quad .
   \label{eq:pi_splittings}
\end{equation}
We map out Autler-Townes spectra for a $\pi$-polarized RF field in Fig. \ref{fig:pipolarized}.
Here, the RF and probe lasers are polarized in $\hat z$, and we rotate the coupling laser polarization slightly off from $\hat z$ towards $\hat x$ to populate all $m_J$ levels. We find excellent agreement between the measured peak locations and the splittings predicted by Eq. \ref{eq:pi_splittings}, with the peak locations matching theory to within one percent. This lineshape is consistent with that predicted by Ref. \cite{cloutman2026rydbergatomicpolarimetryradiofrequency}. Note that at large fields, we observe an avoided crossing between the dressed $D_{5/2}$ and $D_{3/2}$ states. This interaction perturbs the negative detuning side of the spectrum in all of the provided measurements.

While RF circular polarization is less common, it also can be simplified into individual splittings, going as
\begin{equation}
   \Omega = \frac{E_0}{\hbar} \cdot \begin{cases}
        \frac{1}{\sqrt{6}} \mu_0, & m_J = -5/2 \;_{(\rightarrow \text{ -}3/2)}\\
        \frac{1}{\sqrt{10}} \mu_0, & m_J = -3/2\;_{( \rightarrow \text{ -}1/2)}\\
        \frac{1}{\sqrt{20}} \mu_0, & m_J = -1/2\;_{( \rightarrow \;\,1/2)}\\
        \frac{1}{\sqrt{60}} \mu_0, & m_J = \;\;\,1/2\;_{(\rightarrow \;\,3/2)}\\
       0, & m_J = \;\;\,3/2\\
       0,& m_J = \;\;\,5/2\\
   \end{cases}\quad .
   \label{eq:sigma_splittings}
\end{equation}
\section{Conclusion}
We have shown that the structure of Autler–Townes spectra in Rydberg atoms is governed by the polarization of the driving RF field through its coupling of magnetic sublevels. In contrast to the conventional picture of independent $m_J$ transitions, elliptical polarization induces coherent coupling across the full manifold of sublevels, requiring a complete multi-level treatment. By constructing and diagonalizing the corresponding Hamiltonian, we predict polarization-dependent degeneracies that give rise to spectra with two, three, or four resolved peaks.

Using a controlled experimental environment that minimizes RF inhomogeneity and polarization impurities, we directly resolve the magnetic-sublevel structure of the RF-dressed Rydberg states. To our knowledge, this constitutes the first clear experimental resolution of individual $m_J$-dependent features in resonant Rydberg Autler–Townes spectra. The measured peak locations and their evolution with RF ellipticity show excellent agreement with the eigenvalues of the full Hamiltonian. Furthermore, by comparing spectra taken with orthogonal laser polarizations, we extract qualitative information about the angular momentum composition of the dressed states.

These results resolve longstanding discrepancies between theory and experiment in Rydberg Autler–Townes measurements and demonstrate that RF polarization is a critical parameter in interpreting spectroscopic features. More broadly, this work establishes a framework for incorporating magnetic sublevel structure into Rydberg-based sensing, with direct implications for improving the accuracy and traceability of RF electrometry and enabling polarization-resolved measurements in atomic RF sensors.

\section{Methods}
The probe laser had a power of 1.6~$\mu$W in a 300~$\mu$m full-width at half-maximum (FWHM) Gaussian beam, while the probe laser had a power of 70~mW in a 300~$\mu$m FWHM Gaussian beam. To increase the signal-to-noise ratio, the coupling laser power was modulated at 10~kHz with a chopper wheel and the EIT signal was read out from the probe transmission using a lock-in amplifier.
The frequency of the probe laser was locked to an ultra-low expansion cavity while the coupling laser was scanned. The frequency of the coupling laser was calibrated to the spacing of the $D_{3/2}$ and $D_{5/2}$ states. Drift of the coupling laser during data acquisition was accounted for by aligning each trace to the unsplit central peak.

The Cs vapor cell was cylindrical with a 22~mm diameter and 22~mm length. The thickness of the glass walls were around 1~mm. To ensure RF polarization and amplitude uniformity, reflections are minimized with RF-absorbing foam. In addition, the vapor cell was mounted on low density closed cell foam and secured with masking tape in order to minimize scattering of the RF waves.

The background magnetic field was shimmed away with permanent magnets placed on the optics table near the vapor cell.
\section*{Acknowledgements}
The authors thank Samuel Berweger and Benjamin D. Hunt for their insightful comments during the internal review of this manuscript. This research  was supported by NIST under the NIST-on-a-Chip program.  A contribution of the U.S. government, this work is not subject to copyright in the U.S.
The authors declare no conflict of interest.
\section*{Data Availability}
All of the data presented in this paper and used to support the conclusions of this article is available at \cite{MIDAS}.
\bibliography{main}

@article{PhysRevA.109.L021702,
  title = {Zeeman-resolved {Autler-Townes} splitting in {Rydberg} atoms with tunable resonances and a single transition dipole moment},
  author = {Schlossberger, Noah and Rotunno, Andrew P. and Artusio-Glimpse, Alexandra B. and Prajapati, Nikunjkumar and Berweger, Samuel and Shylla, Dangka and Simons, Matthew T. and Holloway, Christopher L.},
  journal = {Phys. Rev. A},
  volume = {109},
  issue = {2},
  pages = {L021702},
  numpages = {6},
  year = {2024},
  month = {Feb},
  publisher = {American Physical Society},
  doi = {10.1103/PhysRevA.109.L021702},
  url = {https://link.aps.org/doi/10.1103/PhysRevA.109.L021702}
}

@article{PhysRevLett.111.063001,
  title = {Atom-Based Vector Microwave Electrometry Using Rubidium {{Rydberg}} Atoms in a Vapor Cell},
  author = {Sedlacek, J. A. and Schwettmann, A. and K\"ubler, H. and Shaffer, J. P.},
  journal = {Phys. Rev. Lett.},
  volume = {111},
  issue = {6},
  pages = {063001},
  numpages = {5},
  year = {2013},
  month = {Aug},
  publisher = {American Physical Society},
  doi = {10.1103/PhysRevLett.111.063001},
  url = {https://link.aps.org/doi/10.1103/PhysRevLett.111.063001}
}

@article{10.1063/5.0173545,
    author = {You, Shu Hang and Cai, Ming Hao and Zhang, Hao An and Song, Zhen Fei and Liu, Hongping},
    title = {RF spectra induced by different polarized microwave},
    journal = {AIP Advances},
    volume = {14},
    number = {1},
    pages = {015245},
    year = {2024},
    month = {01},
    issn = {2158-3226},
    doi = {10.1063/5.0173545},
    url = {https://doi.org/10.1063/5.0173545},
}

@misc{cloutman2026Rydbergatomicpolarimetryradiofrequency,
      title={{Rydberg} atomic polarimetry of radio-frequency fields}, 
      author={Matthew Cloutman and Matthew Chilcott and Alexander Elliott and J. Susanne Otto and Amita B. Deb and Niels Kjærgaard},
      year={2026},
      archivePrefix={arXiv},
      primaryClass={quant-ph},
      url={https://arxiv.org/abs/2503.17997}, 
}

@Article{Schlossberger2024,
author={Schlossberger, Noah
and Prajapati, Nikunjkumar
and Berweger, Samuel
and Rotunno, Andrew P.
and Artusio-Glimpse, Alexandra B.
and Simons, Matthew T.
and Sheikh, Abrar A.
and Norrgard, Eric B.
and Eckel, Stephen P.
and Holloway, Christopher L.},
title={{{Rydberg}} states of alkali atoms in atomic vapour as {SI}-traceable field probes and communications receivers},
journal={Nature Reviews Physics},
year={2024},
month={Oct},
day={01},
volume={6},
number={10},
pages={606-620},
issn={2522-5820},
doi={10.1038/s42254-024-00756-7},
url={https://doi.org/10.1038/s42254-024-00756-7}
}

@article{y8j4-lsbr,
  title = {Population-resolved measurement of an avoided crossing of light-dressed states},
  author = {Schlossberger, Noah and Prajapati, Nikunjkumar and Norrgard, Eric B. and Eckel, Stephen P. and Holloway, Christopher L.},
  journal = {Phys. Rev. A},
  volume = {112},
  issue = {5},
  pages = {053716},
  numpages = {5},
  year = {2025},
  month = {Nov},
  publisher = {American Physical Society},
  doi = {10.1103/y8j4-lsbr},
  url = {https://link.aps.org/doi/10.1103/y8j4-lsbr}
}

@article{10.1063/5.0088532,
    author = {Holloway, Christopher L. and Prajapati, Nikunjkumar and Artusio-Glimpse, Alexandra B. and Berweger, Samuel and Simons, Matthew T. and Kasahara, Yoshiaki and Alù, Andrea and Ziolkowski, Richard W.},
    title = {{Rydberg} atom-based field sensing enhancement using a split-ring resonator},
    journal = {Applied Physics Letters},
    volume = {120},
    number = {20},
    pages = {204001},
    year = {2022},
    month = {05},
    issn = {0003-6951},
    doi = {10.1063/5.0088532},
    url = {https://doi.org/10.1063/5.0088532},
}

@article{10.1063/5.0161213,
    author = {Rotunno, Andrew P. and Holloway, Christopher L. and Prajapati, Nikunjkumar and Berweger, Samuel and Artusio-Glimpse, Alexandra B. and Brown, Roger and Simons, Matthew and Robinson, Amy K. and Kayim, Baran N. and Viray, Michael A. and Jones, Jasmine F. and Sawyer, Brian C. and Wyllie, Robert and Walker, Thad and Ziolkowski, Richard W. and Jefferts, Steven R. and Geibel, Steven and Wheeler, Jonathan and Imhof, Eric},
    title = {Investigating electromagnetically induced transparency spectral lineshape distortion due to non-uniform fields in {Rydberg}-atom electrometry},
    journal = {Journal of Applied Physics},
    volume = {134},
    number = {8},
    pages = {084401},
    year = {2023},
    month = {08},
    issn = {0021-8979},
    doi = {10.1063/5.0161213},
    url = {https://doi.org/10.1063/5.0161213},
}

@book{edmonds1996angular,
  title={Angular Momentum in Quantum Mechanics},
  author={Edmonds, A.R.},
  isbn={9780691025896},
  lccn={77039796},
  series={Investigations in Physics Series},
  url={https://books.google.com/books?id=0BSOg0oHhZ0C},
  year={1996},
  publisher={Princeton University Press}
}

@Article{Sedlacek2012,
author={Sedlacek, Jonathon A.
and Schwettmann, Arne
and K{\"u}bler, Harald
and L{\"o}w, Robert
and Pfau, Tilman
and Shaffer, James P.},
title={Microwave electrometry with {Rydberg} atoms in a vapour cell using bright atomic resonances},
journal={Nature Physics},
year={2012},
month={Nov},
day={01},
volume={8},
number={11},
pages={819-824},
issn={1745-2481},
doi={10.1038/nphys2423},
url={https://doi.org/10.1038/nphys2423}
}

@article{10.1063/5.0285778,
    author = {Talashila, Rajavardhan and Watterson, William J. and Moser, Benjamin L. and Gordon, Joshua A. and Artusio-Glimpse, Alexandra B. and Prajapati, Nikunjkumar and Schlossberger, Noah and Simons, Matthew T. and Holloway, Christopher L.},
    title = {Determining angle of arrival of radio-frequency fields using subwavelength, amplitude-only measurements of standing waves in a {Rydberg} atom sensor},
    journal = {Journal of Applied Physics},
    volume = {138},
    number = {11},
    pages = {114402},
    year = {2025},
    month = {09},
    issn = {0021-8979},
    doi = {10.1063/5.0285778},
    url = {https://doi.org/10.1063/5.0285778},
}

@ARTICLE{9363580,
  author={Anderson, David Alexander and Sapiro, Rachel Elizabeth and Raithel, Georg},
  journal={IEEE Transactions on Antennas and Propagation}, 
  title={A Self-Calibrated {SI}-Traceable {Rydberg} Atom-Based Radio Frequency Electric Field Probe and Measurement Instrument}, 
  year={2021},
  volume={69},
  number={9},
  pages={5931-5941},
  doi={10.1109/TAP.2021.3060540}}

@misc{MIDAS,
    key = {N. Schlossberger, Data associated with ``Resolving magnetic-sublevel structure in {Rydberg Autler-Townes} spectra with
arbitrary {RF} polarization''  National
Institute of Standards and Technology (2026) \href{https://doi.org/10.18434/mds2-4157}{DOI:10.18434/mds2-4157}
}}

@article{PhysRevApplied.4.044015,
  title = {Effect of Vapor-Cell Geometry on {Rydberg}-Atom-Based Measurements of Radio-Frequency Electric Fields},
  author = {Fan, Haoquan and Kumar, Santosh and Sheng, Jiteng and Shaffer, James P. and Holloway, Christopher L. and Gordon, Joshua A.},
  journal = {Phys. Rev. Appl.},
  volume = {4},
  issue = {4},
  pages = {044015},
  numpages = {7},
  year = {2015},
  month = {Oct},
  publisher = {American Physical Society},
  doi = {10.1103/PhysRevApplied.4.044015},
  url = {https://link.aps.org/doi/10.1103/PhysRevApplied.4.044015}
}

@ARTICLE{9374680,
  author={Fancher, Charles T. and Scherer, David R. and John, Marc C. St. and Marlow, Bonnie L. Schmittberger},
  journal={IEEE Transactions on Quantum Engineering}, 
  title={Rydberg Atom Electric Field Sensors for Communications and Sensing}, 
  year={2021},
  volume={2},
  number={},
  pages={1-13},
  doi={10.1109/TQE.2021.3065227}}
\end{document}